\documentclass[aps,prd,floats,nofootinbib,preprintnumbers,superscriptaddress]{revtex4}
\usepackage{epsfig}

\begin{document}

\preprint{Fermilab-Pub-05-029-T}
\preprint{NUHEP-TH/05-03}

\title{Neutrino Mass Hierarchy, Vacuum Oscillations, and Vanishing $|U_{e3}|$}

\author{Andr\'e de Gouv\^ea}
\affiliation{Northwestern University, Department of Physics \& Astronomy, 2145 Sheridan Road, Evanston, IL~60208, USA}
\affiliation{Theoretical Physics Department, Fermilab, PO Box 500, Batavia, IL~60510-0500, USA}

\author{James Jenkins}
\affiliation{Northwestern University, Department of Physics \& Astronomy, 2145 Sheridan Road, Evanston, IL~60208, USA}

\author{Boris Kayser}
\affiliation{Theoretical Physics Department, Fermilab, PO Box 500, Batavia, IL~60510-0500, USA}

\begin{abstract}

Is the relatively isolated member of the neutrino mass spectrum heavier or 
lighter than the two closely-spaced members? This question --- the character 
of the neutrino mass hierarchy --- is of great theoretical interest. All previously identified experiments for addressing 
it via neutrino oscillations require that the currently unknown size of the $U_{e3}$ element of the leptonic mixing matrix (parameterized by the unknown $\theta_{13}$ mixing angle) be sufficiently large, and will utterly fail in the limit $\theta_{13}\to 0$. For this reason, we explore alternative  oscillation approaches that would still succeed even if $\theta_{13}$ vanishes. We identify several alternatives that require neither a nonzero $|U_{e3}|$ nor the presence of significant matter effects. All include multiple percent-level neutrino oscillation measurements, usually involving muon-neutrino (or antineutrino) disappearance and very long baselines. We comment on the degree of promise that these alternative approaches show.

\end{abstract}

\maketitle

\setcounter{equation}{0}
\setcounter{footnote}{0}

\section{Introduction}
\label{sec:intro}

Neutrino experiments have revealed that neutrinos have mass and that, similar to quarks, leptons mix. Neutrino masses were discovered with the observation that neutrinos change flavor as prescribed by the mechanism of mass-induced neutrino oscillations \cite{TASI}.

Three-flavor neutrino oscillations suffice to explain all neutrino data, with the exception of those from LSND \cite{LSND}. The LSND result, however, still awaits confirmation from another experiment and will not be considered here. Nonetheless, even under the conservative assumption that there are no other surprises in the neutrino sector, our understanding of leptons remains incomplete. While we currently know the absolute value of the two neutrino mass-squared differences and two out of four mixing parameters that govern neutrino$\leftrightarrow$neutrino and antineutrino$\leftrightarrow$antineutrino oscillations, we have little or no information regarding\footnote{We adopt the ``standard'' definition of neutrino mass eigenstates (discussed below) and the lepton mixing matrix $U$ \cite{pdg}.} 
\begin{enumerate}
\item the magnitude of the $U_{e3}$-element of the lepton mixing matrix, or the value of $\theta_{13}$;
\item the value of the complex phase $\delta$ which parameterizes CP- (and T-)invariance violation in neutrino$\leftrightarrow$neutrino and antineutrino$\leftrightarrow$antineutrino oscillations;
\item the neutrino mass hierarchy, or the sign of $\Delta m^2_{13}$;
\item the magnitude of the neutrino masses (we only know mass-squared differences and various upper bounds on other combinations of the neutrino masses);
\item whether neutrinos are Majorana or Dirac fermions.
\end{enumerate} 
Items 1, 2, and 3 can be addressed by next-generation neutrino oscillation experiments. Here, we wish to concentrate on item number three and the outstanding question: how can one experimentally determine the neutrino mass hierarchy?

\begin{figure}[ht]
\centerline{\epsfig{width=0.45\textwidth, file=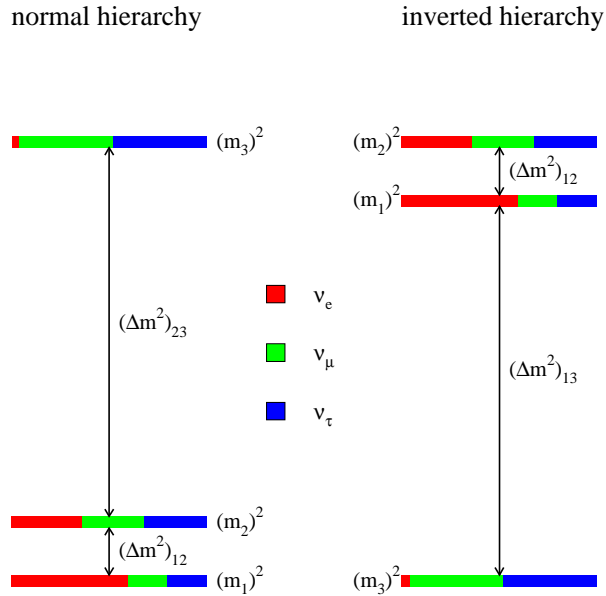}}
\caption{Two distinct neutrino-mass hierarchies that fit all of the current neutrino data, from \cite{TASI}. The color coding (shading) indicates the fraction $|U_{\alpha i}|^2$ of each distinct flavor $\nu_{\alpha}$, $\alpha=e,\mu,\tau$ contained in each mass eigenstate $\nu_i$, $i=1,2,3$. For example, $|U_{e2}|^2$ is equal to the fraction of the $(m_2)^2$ ``bar'' that is painted red (shading labeled as `$\nu_e$').
\label{normal_inverted}}
\end{figure}

The neutrino masses $m_i$, $i=1,2,3$,  are defined in the following convenient way: $m_1^2<m_2^2$, and $0<\Delta m^2_{12}\equiv m^2_2-m^2_1<|\Delta m^2_{13}|$, where $\Delta m^2_{13}\equiv m^2_3-m^2_1$. A positive value of $\Delta m^2_{13}$ implies $m_3^2>m_2^2$ and a so-called normal mass hierarchy, while a negative value of $\Delta m^2_{13}$ implies $m_3^2<m_1^2$ and a so-called inverted mass hierarchy. Examples of an inverted and a normal neutrino mass hierarchy are depicted in Fig.~\ref{normal_inverted}.

The question of whether the neutrino mass hierarchy is normal or inverted is of considerable theoretical interest \cite{neutrino_theory}. For example, one can argue on general grounds that grand unified theories favor a normal neutrino mass hierarchy \cite{albright}. This may be understood qualitatively by noting that grand unified models relate leptons to quarks, and the quark mass hierarchies are normal. An inverted neutrino mass spectrum, on the other hand, would be quite un-quark like, and would probably reflect an underlying global leptonic symmetry (see \cite{neutrino_theory} and references therein) that gives rise to the near degeneracy of $m_1$ and $m_2$.

It is easy to understand, qualitatively, why the current data are unable to distinguish positive from negative $\Delta m^2_{13}$. Almost all information we have on $\Delta m^2_{13}$ comes from $\nu_{\mu}$ disappearance at atmospheric experiments and K2K. These provide information on the $\nu_{\mu}$ survival probability $P_{\mu\mu}$, which is given (in vacuum) by
\begin{equation}
P_{\mu\mu}\simeq1-\sin^22\theta_{23}\sin^2\left(\frac{\Delta m^2_{13}L}{4E}\right),
\label{Pmumu_today}
\end{equation}  
modulo terms proportional to $\Delta m^2_{12}$ or $|U_{e3}|$. These are suppressed because $\Delta m^2_{12}\ll |\Delta m^2_{13}|$ (at the three sigma level, $\Delta m^2_{12}/|\Delta m^2_{13}|<0.06$ \cite{global_anal}), which we know from solar and KamLAND data, and $|U_{e3}|^2\ll 1$ (at the three sigma level, $|U_{e3}|^2<0.051$ \cite{global_anal}), which we know from reactor and atmospheric data. Current experiments are not sensitive to these sub-leading terms, and it is clear that Eq.~(\ref{Pmumu_today}) is only sensitive to $|\Delta m^2_{13}|$. Hence both a normal and an inverted mass hierarchy fit the data equally well.

In this paper, we explore several different oscillation probes of the neutrino mass hierarchy, and discuss under what conditions these can be experimentally accessible. More specifically, we concentrate our efforts on probes other than ``matter-affected 13 oscillations,'' which have been heavily explored in the literature and are defined and very briefly discussed in Sec.~\ref{sec:traditional}. To date, all phenomenological studies of how to determine the neutrino mass hierarchy via neutrino oscillations have, to the best of our knowledge, relied on a non-zero value for $|U_{e3}|$, and almost all appeal to matter effects in the 13 sector. One of the goals of this paper is to argue that neither matter effects nor a nonzero value of $|U_{e3}|$ are required in order to determine the neutrino mass hierarchy via neutrino oscillations --- at least in principle.

There are many reasons for pursuing novel oscillation probes of the neutrino mass hierarchy. One is that we would, ultimately, like to compare distinct determinations of the mass hierarchy. Such a comparison provides a test of whether there is new physics in the lepton sector besides three flavor oscillations. Another (and, perhaps, the most important) reason for pursuing different hierarchy-exploring channels is that, if $\theta_{13}$ turns out to be too small, the traditional approach will not work, so new experimental probes are not only desirable but absolutely necessary!

Our presentation is the following. In Sec.~\ref{sec:traditional}, we briefly review how the mass hierarchy can be revealed if $|U_{e3}|$ is large enough and there are nontrivial matter effects at work. In Sec.~\ref{sec:vacuum}, we remind readers that even in the case of pure vacuum oscillations the mass hierarchy is, in principle, observable. We then review one realistic method for determining the mass hierarchy using only vacuum disappearance channels. In Sec.~\ref{sec:theta13=0}, we study in detail the case $\theta_{13}=0$, and discuss various approaches one might consider in this case to determine the mass hierarchy. In Sec.~\ref{sec:end}, we summarize our results and offer some concluding thoughts. 
A technical presentation of the oscillation probabilities in constant matter for vanishing $\theta_{13}$ is included as an appendix.

Our approach is ``theoretical,'' in the sense that we study the behavior of oscillation probabilities as a function of the neutrino energy $E$ and the baseline $L$ instead of the expected number of observed events for a given neutrino beam, detector type, size, resolution, etc. Nonetheless, our approach allows one to estimate ``how precise'' certain measurements need to be so that the inverted and normal mass hierarchy lead to experimentally distinct results. 

\setcounter{footnote}{0}
\setcounter{equation}{0}
\section{Unveiling The Mass Hierarchy Through Matter Effects in 13 Oscillations}
\label{sec:traditional}

One very promising way of revealing the character of the neutrino mass hierarchy is to probe oscillations which are significantly affected by matter effects in the 13 sector \cite{lbl_pheno,atm_probes,supernovae}. In this section, we very briefly describe one manifestation of this well known phenomenon. Readers are referred to \cite{lbl_pheno,atm_probes,supernovae} for more details. 

By ``matter affected 13 oscillations'' we mean neutrino oscillations that take place under the following two conditions: (i) $|U_{e3}|^2$ is large enough\footnote{The minimum value of $|U_{e3}|^2$ required for unveiling the mass hierarchy via  matter affected 13 oscillations depends on the specific properties of the neutrino source and the detector. For detailed discussions, see \cite{lbl_pheno,atm_probes,supernovae}.} and (ii) the propagating medium contains electron number densities $N_e$ such that the matter potential, $\sqrt{2}G_F N_e$, where $G_F$ is the Fermi constant,  is comparable to the 13 frequency, $|\Delta m^2_{13}|/2E$. Note that, while the latter can be arranged (at least in principle) by choosing an appropriate neutrino beam and propagating medium, the former requires cooperation from Mother Nature. 

Matter affected 13 oscillations can be studied, for example, in accelerator-based, long-baseline searches for $\nu_{\mu}\to\nu_e$ \cite{lbl_pheno}.  In constant matter densities, the $\nu_{\mu}\to\nu_e$ oscillation probability $P_{\mu e}$ is given by
\begin{eqnarray}
&P_{\mu e}\simeq P_{e\mu}\simeq\sin^2\theta_{23}\sin^{2}2\theta^{\rm eff}_{13}\sin^2\left(\frac{\Delta_{13}^{\rm eff}L}{2}\right), \label{P_mue_matter} \\
&\sin^{2}2\theta^{\rm eff}_{13}=\frac{\Delta^2_{13}\sin^22\theta_{13}}{(\Delta^{\rm eff}_{13})^2}, \\
&\Delta_{13}^{\rm eff}=\sqrt{(\Delta_{13}\cos2\theta_{13}-A)^2+\Delta_{13}^2\sin^22\theta_{13}}, \\
&\Delta_{13}=\frac{\Delta m^2_{13}}{2E},
\end{eqnarray}
where $A\equiv\sqrt{2}G_FN_e$ is the matter potential. Eq.~(\ref{P_mue_matter}) is a good approximation as long as $\Delta_{12}$ terms can be ignored.
$\Delta_{13}^{\rm eff}$ is sensitive to the relative sign between $\Delta m^2_{13}\cos 2\theta_{13}$ and $A$. Since $A,\cos2\theta_{13}>0$, a normal neutrino mass hierarchy implies that, for a fixed value of $\theta_{13}$, $P_{\mu e}^{\rm max}$ is {\sl larger} than its vacuum counterpart, $\sin^2\theta_{23}\sin^22\theta_{13}$, while $P_{\bar{\mu}\bar{e}}$ is {\sl smaller}.\footnote{$P_{\bar{\alpha}\bar{\beta}}$ is the $\bar{\nu}_{\alpha}\to\bar{\nu}_{\beta}$ oscillation probability. In matter, antineutrino oscillation probabilities are given by the same expression as the neutrino ones, with $A\to-A$ and $U_{\alpha i}\to U_{\alpha i}^*$.} The opposite behavior is expected for an inverted hierarchy. 

While $P_{\mu e}$ is sensitive to the mass hierarchy, measuring it alone at approximately fixed values of $L$ and $E$ is often not enough to experimentally determine the mass hierarchy. This is due to a well known $\theta_{13}$--mass hierarchy degeneracy, which can be understood in the following way. If one only measures $\sin^{2}2\theta^{\rm eff}_{13}$, for every normal hierarchy characterized by $(\Delta m^2_{13},\theta_{13})=(\Delta m^{2+}_{13},\theta^+_{13})$, there will be an inverted one with $(\Delta m^2_{13},\theta_{13})=(\Delta m^{2-}_{13},\theta^-_{13})$ that yields exactly the same value of $\sin^{2}2\theta^{\rm eff}_{13}$. Note that, as long as $\Delta m^{2+}_{13}\sim -\Delta m^{2-}_{13}$,\footnote{This is required if we wish our two candidate hierarchies to fit all currently available atmospheric neutrino data.} the degeneracy is dominated by the relative size of $\theta_{13}^+$ and $\theta_{13}^-$ and the sign of $\Delta m^2_{13}$. Qualitatively, $\sin^{2}2\theta^{\rm eff}_{13}$ can be chosen the same for a normal and an inverted mass hierarchy if one chooses $\theta_{13}^+$ to be less than $\theta_{13}^-$.

The degeneracy described above can be broken by more experimental information. Simple ways of determining the mass hierarchy include comparing $P_{\mu e}$ and $P_{\bar{\mu}\bar{e}}$ at $\Delta_{13} L\sim \pi$ (which can be done with one detector and both muon-neutrino and muon-antineutrino beams), or comparing measurements of $P_{\mu e}$ at $\Delta_{13} L\sim \pi$ for different values of $L$ (which requires two different detectors, but only muon-neutrino beams). 
Another way is to independently measure $\sin^22\theta_{13}$ precisely enough such that one can statistically distinguish between $\theta_{13}=\theta^+_{13}$ and $\theta_{13}=\theta_{13}^-$.

Other probes of the neutrino mass hierarchy that depend on matter effects and a nonzero $|U_{e3}|$ include precise atmospheric neutrino studies \cite{atm_probes} and  measurements of the neutrino flux emitted during a nearby supernova explosion \cite{supernovae}. 

The procedures referred to above do not take advantage of the fact that, in the case of three neutrino flavors, neutrino oscillations are governed by two distinct oscillation frequencies. Indeed, extracting the neutrino mass hierarchy via matter affected 13 oscillations works perfectly well even in the limit $\Delta m^2_{12}\to 0$, when solar effects become utterly unobservable \cite{lbl_low}. All procedures referred to above, however, rely fundamentally on a nonzero value of $|U_{e3}|$ and neutrino oscillations in matter. In the next two sections, we argue that neither the former nor the latter is required in order to determine the mass hierarchy through neutrino oscillations --- at least in principle.

To conclude this section, it is illustrative to mention  that the ``solar mass-hierarchy'' was revealed by measuring how matter effects modify oscillations in the 12 sector. Since $\Delta m^2_{12}$ is chosen positive-definite,  the ``solar mass-hierarchy'' question is the following: is the $\nu_e$ flavor eigenstate ``predominantly $\nu_1$'' or ``predominantly $\nu_2$,'' {\it i.e.}, is $|U_{e1}|^2$ larger or smaller than $|U_{e2}|^2$?\footnote{Since $|U_{e1}|^2/|U_{e2}|^2\equiv\tan^2\theta_{12}$, this is equivalent to inquiring about the sign of $\cos2\theta_{12}$, or whether the solar angle is in the light or the dark-side of the parameter space \cite{darkside}.} The electron-type neutrino survival probability in vacuum, assuming that $\Delta m^2_{13,23}$ effects are averaged out, is incapable of addressing this question. Under these circumstances, $P_{ee}$ is only a function of $|U_{e1}|^2|U_{e2}|^2$ and hence invariant under $|U_{e1}|^2\leftrightarrow|U_{e2}|^2$. Matter effects break the degeneracy, such that, for $^8$B solar neutrinos, $P_{ee}\simeq |U_{e2}|^2$. Current data reveal that $|U_{e2}|^2<0.5$ (and hence $|U_{e1}|^2\simeq 1-|U_{e2}|^2>|U_{e2}|^2$), at more than the five sigma level. One reason we have been able to disentangle the solar mass hierarchy is that $\theta_{12}$ is ``different enough'' from  $0,\pi/2$. Had that not been the case, we would have been unable to determine whether $\nu_e\sim\nu_1$ ($\theta_{12}=0$) or $\nu_e\sim\nu_2$ ($\theta_{12}=\pi/2$) --- clearly a physical question! --- via matter affected 12 oscillations.

\setcounter{footnote}{0}
\setcounter{equation}{0}
\section{Unveiling the Hierarchy Through Vacuum Oscillations}
\label{sec:vacuum}

In the case of only two neutrino flavors, the neutrino mass hierarchy \cite{darkside} is simply unobservable when it comes to studying vacuum flavor oscillations. This is easy to see. With only two flavors, $P_{\alpha\beta}=1-P_{\alpha\alpha}=\sin^22\theta\sin^2(\Delta m^2L/4E)$, and there is simply no sensitivity to the sign of $\Delta m^2$ or the sign of $\cos2\theta$. This does not, of course, mean that the mass hierarchy is not physical --- it is just not observable if one measures only vacuum oscillation probabilities. 

In the case of three or more flavors, this is no longer the case. Pure vacuum oscillations can, in general, determine whether the neutrino mass hierarchy, as defined in Sec.~\ref{sec:intro}, is normal or inverted. In this section we explicitly present how this can be done in principle, and then review how it could be done in practice by comparing accelerator $\nu_{\mu}$ and reactor $\bar{\nu}_e$ disappearance searches.

\subsection{How To Compare a Normal and an Inverted Mass-Hierarchy}

In vacuum, the survival probability $P_{\alpha\alpha}$  can be written as
\begin{equation}
P_{\alpha\alpha}=1-4|U_{\alpha1}|^2|U_{\alpha2}|^2\sin^2\left(\frac{\Delta_{12}L}{2}\right)-
4|U_{\alpha1}|^2|U_{\alpha3}|^2\sin^2\left(\frac{\Delta_{13}L}{2}\right)-
4|U_{\alpha2}|^2|U_{\alpha3}|^2\sin^2\left(\frac{\Delta_{23}L}{2}\right),
\label{eq:p_vac}
\end{equation}
where $\Delta_{ij}\equiv\Delta m^2_{ij}/2E$. Note that $\Delta_{23}=\Delta_{13}-\Delta_{12}$. Eq.~(\ref{eq:p_vac}) is valid, of course, independent of the mass hierarchy.

We wish to probe whether normal and inverted mass hierarchies necessarily lead to different values of $P_{\alpha\alpha}$. We will concentrate our efforts on comparing inverted and normal hierarchies which are capable of fitting the current neutrino data. More specifically,  we will only consider normal and inverted mass hierarchies that share the same value of $\Delta m^2_{12}$ and the same mixing matrix. These constraints clearly satisfy the requirement above and substantially simplify our presentation. Relaxing these constraints would render our analyses much more opaque, and would not modify qualitatively our results. 

Hence, we address the following question: for a given choice of $\Delta m^2_{13}=\Delta m^{2+}_{13}$ corresponding to a normal mass hierarchy ($\Delta m^{2+}_{13}>\Delta m^{2}_{12}>0$), is there an inverted mass spectrum characterized by the same mixing matrix and $\Delta m^2_{12}$ but a different $\Delta m^2_{13}=\Delta m^{2-}_{13}$ ($-\Delta m^{2-}_{13}>\Delta m^{2}_{12}>0$) that yields exactly the same survival probability $P_{\alpha\alpha}$, for all values of $L/E$?

Define $x$ by $\Delta m^{2-}_{13}=-\Delta m^{2+}_{13}+x$. Note that $x$ is restricted to be small enough so that $\Delta m^{2-}_{13}$ is indeed a negative quantity.\footnote{As a matter of fact, the magnitude of $x$ should be small enough so that both $\Delta m^{2+}_{13}$ and $|\Delta m^{2-}_{13}|$ agree with the currently allowed range of values for the ``atmospheric'' mass-squared difference.} In practice, $x$ turns out to be of order $\Delta m^2_{12}$, as we will see shortly. The difference between $P_{\alpha\alpha}(\Delta m^{2+}_{13})$ and $P_{\alpha\alpha}(\Delta m^{2-}_{13})$, $P^+_{\alpha\alpha}-P^-_{\alpha\alpha}$, is
\begin{eqnarray}
P^+_{\alpha\alpha}-P^-_{\alpha\alpha}&=& -4
|U_{\alpha3}|^2\left\{|U_{\alpha1}|^2\left[\sin^2\left(\frac{\Delta_{13}L}{2}\right)-\sin^2\left(\frac{(\Delta_{13}-X)L}{2}\right)\right] \right.  \nonumber\\
&+&\left.|U_{\alpha2}|^2\left[\sin^2\left(\frac{(\Delta_{13}-\Delta_{12})L}{2}\right)-\sin^2\left(\frac{(\Delta_{13}+\Delta_{12}-X)L}{2}\right)\right]\right\},
\label{P_vac_diff}
\end{eqnarray}
where $X\equiv x/2E$ and, for simplicity, we have dropped the `$+$' superscript in $\Delta m^{2+}_{13}$. There is, in general, no constant value of $x$ that renders Eq.~(\ref{P_vac_diff}) exactly zero for all $L$.\footnote{Keep in mind that $x$ is restricted to be small, such that $\Delta m^{2-}_{13}<0$. $P^+-P^-=0$ is guaranteed to be true for $x=2\Delta m^2_{13}$, independent of whether the oscillations occur in matter or in vacuum. This is trivial to understand, since $x=2\Delta m^2_{13}$ renders the two mass patterns in question identical --- such a large $x$ value ``inverts the inverted hierarchy.''} The difference in the first square-bracket requires $x=0$, while the difference in the second one calls for $x=2\Delta m^2_{12}$. Curiously enough, there is a special solution in the case $|U_{\alpha1}|^2=|U_{\alpha2}|^2$: $x=\Delta m^2_{12}$.

Qualitatively, it is easy to understand what is happening. The distinction between a normal and an inverted mass hierarchy is, as far as vacuum oscillations are concerned, the following. For a normal mass hierarchy, $|\Delta m^2_{13}|>|\Delta m^2_{23}|$. For an inverted one, the situation is reversed: $|\Delta m^2_{13}|<|\Delta m^2_{23}|$. One can choose, as is depicted in Fig.~\ref{normal_inverted}, all normal and inverted mass hierarchy frequencies ($|\Delta_{ij}|$) to agree by equating $|\Delta m^{2+}_{13}|=|\Delta m^{2-}_{23}|$ and $|\Delta m^{2-}_{13}|=|\Delta m^{2+}_{23}|$ --- equivalent to  $x=\Delta m^2_{12}$. In this case, however, $P^+_{\alpha\alpha}\neq P^-_{\alpha\alpha}$, because the $\Delta_{13}$ and $\Delta_{23}$ amplitudes are, in general, different (except in the special case $|U_{\alpha1}|^2=|U_{\alpha2}|^2$). 

We conclude that it should be possible to distinguish between the two hierarchies by studying neutrino oscillations in vacuum. The key requirement  is the ability to tell $\Delta m^2_{13}$ from $\Delta m^2_{23}$ or, in other words, simultaneous sensitivity to $\Delta m^2_{12}$ {\rm and} $\Delta m^2_{13}$ effects. All experiments performed to date fail to have this simultaneous sensitivity for different reasons. ``$\Delta m^2_{12}$ experiments,'' like KamLAND, lack sensitivity to $\Delta m^2_{13}$ effects, which ``average out''.\footnote{On top of that, $|U_{e3}|$ turns out to be too small.} ``$\Delta m^2_{13}$ experiments,'' like K2K, lack sensitivity to $\Delta m^2_{12}$ effects, which are suppressed by a small $\Delta_{12}L$.

From the discussion above, it seems that an experiment capable of measuring $\Delta m^2_{13}$ with an uncertainty $\Delta(\Delta m^2_{13})<\Delta m^2_{12}$ should be able to tell a normal from an inverted mass hierarchy. However, it turns out that this is not automatically the case. 

Assume one is probing $P_{\alpha\alpha}$ with neutrino energies and baselines such that $\Delta_{13}L\sim 1$, as is the case for proposed next-generation accelerator and reactor neutrino experiments \cite{next-generation}. Under these circumstances, $\Delta_{12}L\ll 1$ and
\begin{equation}
P^+_{\alpha\alpha}-P^-_{\alpha\alpha}=-2L|U_{\alpha3}|^2\sin\left(\Delta_{13}L\right)|\left[|U_{\alpha1}|^2X 
+|U_{\alpha2}|^2\left(-\Delta_{12} +(X-\Delta_{12}\right)\right] 
+O\left((\Delta_{12} L)^2,(XL)^2\right).
\label{P_small_diff}
\end{equation}
We are interested in values of $x$ such that $XL\ll 1$, and have made use of the fact that
\begin{equation}
\sin^2(\theta-\epsilon)=\sin^2\theta-\epsilon\sin2\theta+O(\epsilon^2).
\end{equation}
One can easily find $x$ such that, neglecting $O((\Delta_{12}L)^2,(XL)^2)$ terms, $P^+_{\alpha\alpha}-P^-_{\alpha\alpha}=0$:
\begin{equation}
x=\frac{2|U_{\alpha2}|^2}{|U_{\alpha1}|^2+|U_{\alpha2}|^2}\Delta m^2_{12},
\label{eq:xaa}
\end{equation}
for $|U_{\alpha3}|^2\neq 0$ and ignoring the possibility that $\sin(\Delta_{13}L)=0$.

This  means that if one probes, say, $P_{\mu\mu}$ in vacuum at values of $L$ and $E$ such that $\Delta_{12}L\ll 1$ and measures, assuming a normal mass-hierarchy, $\Delta m^{2+}_{13}=(2.13\pm 0.04)\times 10^{-3}$~eV$^2$ (a 2\% measurement of the ``atmospheric'' mass-squared difference), there will be a statistically equivalent solution for an inverted mass hierarchy and $\Delta m^{2-}_{13}=-(2.02\pm 0.04)\times 10^{-3}$~eV$^2$ if $x_{\mu\mu}=0.11\times 10^{-3}$~eV$^2$.\footnote{We define $x_{\alpha\alpha}$, $\alpha=e,\mu,\tau$ as the value of $x$ required to render the $\nu_{\alpha}$ survival probability for an inverted and normal hierarchy indistinguishable in a range of values of $L$ and $E$. Similarly, we define $x_{\alpha\beta}$ as the equivalent quantity for $\nu_{\alpha}\to\nu_{\beta}$ transitions ($\alpha,\beta=e,\mu,\tau$).} Today, $x_{\mu\mu}$ is constrained to be
\begin{equation}
x_{\mu\mu}\equiv\frac{2|U_{\mu2}|^2}{1-|U_{\mu3}|^2}\Delta m^2_{12}\in[0.044,0.17]\times 10^{-3}~\rm eV^2,
\label{eq:xmu}
\end{equation}  
taking three sigma upper and lower bounds for the oscillation parameters \cite{global_anal}.

Appearance channels also contain a similar normal--inverted degeneracy. If one were to probe $\nu_{\alpha}\to\nu_{\beta}$ in a setup where $\Delta_{12}L\ll 1$, a normal and inverted mass-hierarchy would yield the same oscillation probability $P_{\alpha\beta}$ (up to $O((\Delta_{12}L)^2)$ effects) if 
\begin{equation}
x_{\alpha\beta}=-2\cos\xi\left|\frac{U_{\alpha2}U_{\beta2}}{U_{\alpha3}U_{\beta3}}\right|\Delta m^2_{12},
\label{eq:xab}
\end{equation}
where $\xi=\arg[U_{\alpha3}U^*_{\beta3}U^*_{\alpha2}U_{\beta2}]$.

\subsection{Comparing $P(\bar{\nu}_e\to\bar{\nu}_e)$ to $P(\nu_{\mu}\to\nu_{\mu})$}

One way to eliminate the degeneracy presented above is to compare two distinct oscillation channels, since the value of $x$ that renders inverted and normal hierarchies indistinguishable depends on the parent and daughter neutrino flavors. For concreteness, we will discuss under what conditions one can determine the mass-hierarchy by studying $\bar{\nu}_e$ and $\nu_{\mu}$ disappearance (at, say, a reactor neutrino experiment and a long-baseline conventional neutrino beam setup, respectively). 
One variant of this procedure has already been explored (using a different ``language'') in the literature \cite{petcov_piai} and is, to the best of our knowledge, the only oscillation probe of the neutrino mass hierarchy identified so far that does not rely significantly on matter effects. 
In the case of a reactor neutrino experiment, the vacuum expression for the survival probability is a very good approximation. For accelerator-based experiments, this is only true for low-energy neutrinos ($E<$~few hundred MeV) assuming that the neutrinos are propagating inside the Earth's crust. Some consequences of nontrivial matter effects are discussed in Secs.~\ref{sec:traditional} (for nonzero $|U_{e3}|$) and \ref{sec:theta13=0} (for $|U_{e3}|=0$).

\begin{figure}
{\centering
\epsfig{figure=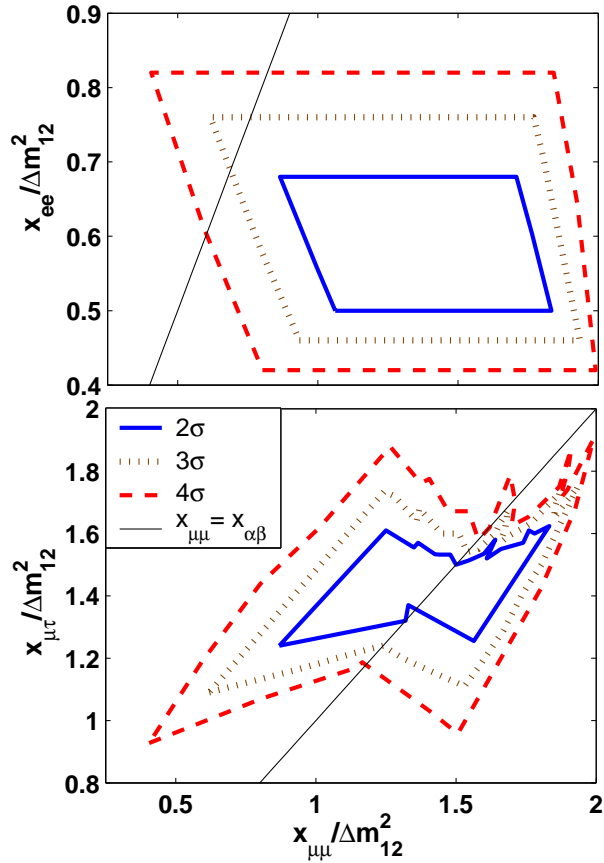,width=0.45\textwidth}
}
\caption{Currently allowed values of $x_{\mu\mu}\times x_{ee}$ (top) and $x_{\mu\mu}\times x_{\mu\tau}$(bottom) (in units of $\Delta m^2_{12}$) taking two (solid), three (dotted) and four (dashed) sigma allowed values of all oscillation parameters, according to \cite{global_anal}. See text for details. The thin solid line corresponds to $x_{ee},x_{\mu\tau}=x_{\mu\mu}$.}
\label{xmu_xx}
\end{figure}

In the $\nu_{\mu}$ disappearance experiment, the two hierarchies can be made compatible by choosing $x=x_{\mu\mu}$, as defined in Eq.~(\ref{eq:xmu}). On the other hand, for the reactor experiment, the two hierarchies will yield identical results if $x=x_{ee}$, given by
\begin{equation}
x_{ee}=\frac{2|U_{e2}|^2}{1-|U_{e3}|^2}\Delta m^2_{12}=2\Delta m^2_{12}\sin^2\theta_{12}\in [0.033,0.068]\times 10^{-3}~\rm eV^2,
\end{equation}
taking current three sigma upper and lower bounds for $\sin^2\theta_{12}$ and $\Delta m^2_{12}$. As long as $x_{\mu\mu}\neq x_{ee}$, the ``wrong hierarchy'' measurements of $\Delta m^2_{13}$ obtained via the electron-type antineutrino and muon-type neutrino disappearance analyses will disagree, thus breaking the degeneracy. Figure \ref{xmu_xx}(top) depicts the currently allowed values of $x_{\mu\mu}$ and $x_{ee}$, in units of $\Delta m^2_{12}$, if all oscillating parameters are allowed to vary within their two, three, and four sigma allowed ranges \cite{global_anal}. 

In order to realistically determine the mass hierarchy, both the reactor and accelerator experiments are required to measure the atmospheric mass-squared difference with great precision: $\Delta(\Delta m^2_{13})< |x_{\mu\mu}-x_{ee}|$. Given what is currently known of the oscillation parameters, these could conspire in such a way that $|x_{\mu\mu}-x_{ee}|\ll\Delta m^2_{12}$, as can be clearly seen in Fig.~\ref{xmu_xx}(top). Improved knowledge of the oscillation parameters (especially $\theta_{12}$ and $\theta_{13}$) will help determine how challenging this proposition really is. 

While a 2\% measurement of $\Delta m^2_{13}$ in an accelerator $\nu_{\mu}$ disappearance experiment does not seem too far-fetched, as precise a measurement in a reactor experiment seems too optimistic. ``Beta beams,'' for example, may serve as a more advantageous alternative capable of precisely studying $\nu_e\to\nu_e$ transitions in vacuum \cite{beta_beam}. We are aware of an ongoing in-depth analysis of how to determine the neutrino mass hierarchy via a comparison of $\nu_{\mu}\to\nu_{\mu}$ and $\nu_{e}\to\nu_e$ ($\bar{\nu}_{e}\to\bar{\nu}_e$), to appear shortly \cite{Parke}.

Similar options which are potentially useful include comparing $\nu_{\mu}\to\nu_{\mu}$ to $\nu_{\mu}\to\nu_{\tau}$ or $\nu_{\mu}\to\nu_e$. The currently allowed values of $x_{\mu\mu}$ and $x_{\mu\tau}$ are depicted in Fig.~\ref{xmu_xx}(bottom). Needless to say, precisely measuring $\Delta m^2_{13}$ in an electron or tau appearance channel seems far from realistic. Furthermore, as far as tau appearance experiments are concerned, neutrino energies are required to be significantly larger than 1~GeV, and the pure vacuum approximations we have been discussing so far simply do not apply in the case of an Earth-based accelerator experiment. We will briefly comment on the $\nu_{\mu}\to\nu_{\tau}$ channel in the next section. 

Comparing the inverted--normal hierarchy degeneracy in $\nu_{\mu}\to\nu_{\mu}$ and $\nu_{\mu}\to\nu_{e}$ oscillations requires more care. The value of $x_{\mu e}$ obtained from Eq.~(\ref{eq:xab}) diverges as $|U_{e3}|^{-1}$ as $|U_{e3}|\to 0$, and one may be tempted to conclude that $x_{\mu e}$ could end up much larger than $x_{\mu\mu}$, rendering this comparison very sensitive to the mass hierarchy. This is not necessarily the correct conclusion for a few reasons. One is that, for small values of  $|U_{e3}|$, the $\Delta m^2_{13}$ dependency of the $\nu_{\mu}\to\nu_e$ oscillation probability disappears, and at some point all sensitivity to ``atmospheric'' oscillations goes away --- $x_{\mu e}$ may be large, but not experimentally accessible! Another is that, in the limit $x\gg \Delta m^2_{12}$, Eq.~(\ref{eq:xab}) need not apply, since it was derived under the assumption that $XL\ll 1$.

\setcounter{footnote}{0}
\setcounter{equation}{0}
\section{The Mass Hierarchy and Vanishing $|U_{e3}|$}
\label{sec:theta13=0}

We emphasized in Sec.~\ref{sec:traditional} that when $|U_{e3}|\to 0$ the traditional approach to determining the mass hierarchy through matter effects in the 13 sector fails, and new exploration channels are not only desirable but necessary. Furthermore, the approach studied in the previous section also fails in the limit $|U_{e3}|\to 0$. One reason is simply that, in this case, oscillations involving electron neutrinos do not depend on $\Delta m^2_{13}$ (including its sign).\footnote{One concrete manifestation of this can be found in Appendix~\ref{app:tech}.} Hence, in the limit $|U_{e3}|\to0$, we are forced to explore the neutrino mass hierarchy through $P_{\mu\mu}$, $P_{\mu\tau}$ and $P_{\tau\tau}$. 

For vanishing $\theta_{13}$, neutrino oscillation experiments can still distinguish, in principle, an inverted mass hierarchy from a normal one. For example, in vacuum, $P_{\mu\mu}^+-P_{\mu\mu}^-$ (see Eq.~(\ref{P_vac_diff})) still does not vanish for all  $L/E$ for any choice of $x$ if $\theta_{13}=0$, unless $\theta_{12}=\pi/4$ (maximal solar mixing, experimentally excluded). The same is true of $P_{\mu\tau}^+-P_{\mu\tau}^-$. It is, however, curious that no new information is obtained if one compares the muon disappearance, tau disappearance, and tau appearance channels (for the same value of $L$ and $E$).  This can be seen in a variety of ways and boils down to the fact that, for $U_{e3}=0$, all three oscillation channels change ``in the same way'' when one compares a normal to an inverted mass hierarchy. For example, in the limit $|U_{e3}|=0$ and $\Delta_{12}L\ll 1$,
\begin{equation}
x_{\mu\mu}=x_{\tau\tau}=x_{\mu\tau}=2\Delta m^2_{12}\cos^2\theta_{12},
\label{eq:mumu=tautau}
\end{equation}
using Eqs.~(\ref{eq:xaa},\ref{eq:xab}). The same behavior is observed in matter or outside the limit $\Delta_{12}L\ll 1$. For details, see Appendix~\ref{app:tech}. It suffices, therefore, to concentrate one's efforts on the muon disappearance channel, which is what we do in this section. Throughout, we will assume $|U_{e3}|=0$, unless otherwise noted.

\subsection{Comparing Values of $P(\nu_{\mu}\to\nu_{\mu})$ in Vacuum at Different $L/E$}

In Sec.~\ref{sec:vacuum}, we argued that as long as experiments are performed under conditions such that $\Delta_{12}L\ll 1$, for every normal hierarchy choice of mass-squared differences there is an inverted hierarchy one that yields the same oscillation probability (up to $(\Delta_{12}L)^2$ corrections): $\Delta m^{2-}_{13}=-\Delta m^{2+}_{13}+x$, where $x$ is given by either Eq.~(\ref{eq:xaa}) or Eq.~(\ref{eq:xab}). In order to break this degeneracy, it is clear that one needs to explore setups where $\Delta_{12}L\sim1$.

Values of $\Delta_{12}L\sim 1$ imply very long baselines or very small neutrino energies. Given that we are considering experiments in which a muon is created, we need to steer clear from very low energies, and are forced to look at long baselines. Under these circumstances, it is quite likely that we will have to deal with relevant matter effects (short of considering experiments that shoot neutrino beams into space). Nonetheless, we first warm-up by discussing how one should go about disentangling the mass-hierarchy with muon disappearance in pure vacuum oscillations, and turn to matter effects in the next subsection.  

\begin{figure}
{\centering
\epsfig{figure=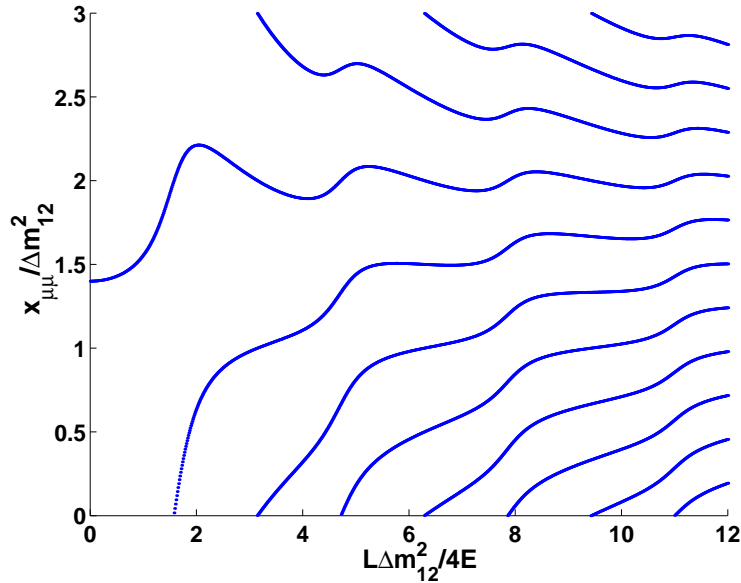,width=0.55\textwidth}
}
\caption{Values of $x_{\mu\mu}/\Delta m^2_{12}$ that render $P_{\mu\mu}^+(\Delta_{12}L/2)=P_{\mu\mu}^-(\Delta_{12}L/2)$ as a function of $\Delta_{12}L/2$, for $\cos2\theta_{12}=0.4$ ($\sin^2\theta_{12}=0.3$). See text for details. The plot range is restricted to $x_{\mu\mu}$ values such that $|x_{\mu\mu}|$ is of order $\Delta m^2_{12}$.}
\label{Fig:xvac}
\end{figure}

We start by making the following point: for a fixed $L$ and $E$ there are always values of $x$ that renders $P^+-P^-=0$. Explicitly, $P_{\mu\mu}^+-P_{\mu\mu}^-=0$ translates (after a long series of trigonometric manipulations) into
\begin{equation}
\sin\left(\frac{(2\Delta_{13}+X)L}{2}\right)\left[\tan\left(\frac{(\Delta_{12}-X)L}{2}\right)+(\cos2\theta_{12})\tan\left(\frac{\Delta_{12}L}{2}\right)\right]=0.
\end{equation}
The values of $x$ that satisfy the trigonometric equation above are $x=-2\Delta m^2_{13}+4n\pi\left(\frac{E}{L}\right)$, integer $n$, and 
\begin{equation}
x=\Delta m^2_{12}+\frac{4E}{L}\arctan\left(\cos2\theta_{12}\tan\left(\frac{\Delta_{12}L}{2}\right)\right)+4n\pi\left(\frac{E}{L}\right),
\label{eq:x(L/E)}
\end{equation}
for integer values of $n$ and the appropriate definition of the inverse tangent function. The first set of solutions violates the restrictions we impose on the allowed values of $x$ (as already pointed out, $x=-2\Delta m^2_{13}$ transforms an inverted hierarchy into a normal one) and will be ignored. Values of $x$ that satisfy Eq.~(\ref{eq:x(L/E)}) are depicted in Fig.~\ref{Fig:xvac} as a function of $\Delta_{12}L/2$, for $\cos2\theta_{12}=0.4$. As expected (see Eq.~(\ref{eq:mumu=tautau})), for $\Delta_{12}L\ll 1$, $x=2\cos^2\theta_{12}\Delta m^2_{12}$ for $n=0$.

Figure~\ref{Fig:xvac} reveals that, for a fixed $L/E$, there are values of $x$ of order $\Delta m^2_{12}$ that render $P^+=P^-$, even when $\Delta_{12}L=O(1)$. Indeed, for large enough values of $\Delta_{12}L$, there are several (discreet) values of $x$ of order $\Delta m^2_{12}$ that do the trick.\footnote{As $\Delta_{12}L\to 0$, all $x$ values that obey Eq.~(\ref{eq:x(L/E)}) for $n\neq 0$ diverge, and only the $n=0$ solution is acceptable.} On the other hand, there is no constant value of $x$ capable of solving $P^+=P^-$ for all $L/E$, as already extensively discussed. It is, therefore, clear that while a normal and an inverted mass hierarchy lead to distinct $P_{\mu\mu}$, if $\nu_{\mu}\to\nu_{\mu}$ processes are studied around a fixed value of $L/E$, it is not possible to tell one mass hierarchy from the other. In order to break the degeneracy, more experimental information is required.

For example, if one were to perform a very precise ``short baseline'' experiment characterized by small values of  $\Delta_{12}L$, one would find two degenerate solutions for the ``atmospheric'' mass-squared difference: $\Delta m^{2+}_{13}$ and $-\Delta m^{2+}_{13}+x_{\rm sbl}$. On the other hand, a ``long baseline'' narrow-band-beam experiment with typical energy and baseline such that $\Delta_{12}L\gtrsim 1$, would measure, assuming that the neutrino mass hierarchy is indeed normal,  $\Delta m^{2+}_{13}$ and $-\Delta m^{2+}_{13}+x_{\rm lbl}$. The mass hierarchy would be determined by combining both data sets as long as the precision $\Delta(\Delta m^2_{13})$ with which {\sl both} experiments measure $\Delta m^2_{13}$ is better than $|x_{\rm sbl}-x_{\rm lbl}|$. As an example, if $\Delta_{12}L\sim 4$ at the ``long-baseline'' site,  $|x_{\rm sbl}-x_{\rm lbl}|\sim 0.8\Delta m^2_{12}$. 

Fig.~\ref{Fig:xvac} can also be used to estimate how precise a broad-band-beam experiment in vacuum needs to be in order to establish the mass hierarchy. If one takes this experiment's data and ``bins'' it in $L/E$,  each $L/E$ bin (or $\Delta_{12}L$ bin) has to be capable of measuring $\Delta m^2_{13}$ with $\Delta(\Delta m^2_{13})\lesssim \Delta m^2_{12}$.\footnote{A narrow $\Delta_{12}L$ bin will contain several $\Delta m^2_{13}$ ``wavelengths,'' which allows one to, in principle, perform a measurement of $\Delta m^2_{13}$. We will return to this issue with more details in the next subsection.} The same is true, of course, of experiments with variable baselines, most notably atmospheric neutrino experiments. We return to these issues and make a more realistic assessment of the situation in the next subsection.

\subsection{Comparing Values of $P(\nu_{\mu}\to\nu_{\mu})$ in Matter at Different $L$ and $E$}

When $|U_{e3}|=0$, the importance of matter effects is estimated by comparing $\Delta_{12}$ with the matter potential $A$.  Numerically
\begin{equation}
\frac{\Delta_{12}}{A}=3.5\left(\frac{\Delta m^2_{12}}{8\times 10^{-5}~\rm eV^2}\right)\left(\frac{100~\rm MeV}{E}\right)\left(\frac{\rho}{3 \rm g/cm^3}\right)^{-1}.
\end{equation}
We are interested in neutrino energies larger than 100~MeV, so that $\nu_{\mu}X\to\mu X^{\prime}$ charged current processes are kinematically allowed, and typical Earth densities, which are around 2--5~g/cm$^3$. 

As shown in Appendix~\ref{app:tech}, $P_{\mu\mu}$ in  matter can be written as $P_{\mu\mu}$ in vacuum with $\Delta_{ij}$ and $\theta_{12}$ replaced by effective quantities. Matter effects will modify neutrino oscillations (and our ability to determine the mass hierarchy) in two ways:
\begin{itemize}
\item $\Delta_{13}^{m}-\Delta_{23}^m=\Delta_{12}^m$, so that the difference between the effective 13 and 23 oscillation frequencies will be different than $\Delta_{12}$. In the case of vacuum oscillations, this difference was recognized as the quantity that determines how precisely one needs to measure $\Delta_{13}$ effects in order to establish the mass hierarchy;
\item $\sin2\theta_{12}^m=(\Delta_{12}/\Delta_{12}^m)\sin2\theta_{12}$, so that the effective solar angle is different from its vacuum counterpart. In the case of vacuum oscillations, $\tan^2\theta_{12}$ measures the ratio of the amplitudes associated with 13 and 23 oscillations, and plays an important role in determining the value of $x$ which renders $P_{\mu\mu}^+=P_{\mu\mu}^-$ for a fixed $L/E$.
\end{itemize}

Two limiting cases are particularly easy to describe. One is the limit of strong matter effects: $|A|\gg\Delta_{12}$. Under these circumstances, $\Delta_{12}^m\gg \Delta_{12}$, and $\sin^22\theta_{12}^m\to 0$. Hence, 
\begin{equation}
P_{\mu\mu}^{A\gg\Delta_{12}}\simeq 1-\sin^22\theta_{23}\sin^2\left(\frac{(\Delta_{13}-\Delta_{12}\cos^2\theta_{12})L}{2}\right),
\label{Pmumu_largeA}
\end{equation}
in which case $x=2\Delta m^2_{12}\cos^2\theta_{12}$ renders $P_{\mu\mu}^+=P_{\mu\mu}^-$. This can be understood in the following way. In the case of very strong matter effects and neutrino (antineutrino) oscillations, $\sin^2\theta_{12}^m\to 1$ ($\cos^2\theta_{12}^m\to 1$), so that $\nu_{\mu}\leftrightarrow\nu_{\mu}$ ($\bar{\nu}_{\mu}\leftrightarrow\bar{\nu}_{\mu}$) oscillations are governed by only one oscillation frequency \cite{strong_matter}.\footnote{$\Delta_{13}-\Delta_{12}\cos^2\theta_{12}=\Delta_{13}\sin^2\theta_{12}+\Delta_{23}\cos^2\theta_{12}$. Hence, high energy ($E>$~few GeV) $\nu_{\mu}\leftrightarrow\nu_{\mu}$ oscillations inside the Earth probe neither $\Delta m^2_{13}$ nor $\Delta m^2_{23}$, but the solar-mixing weighted average of the two. This agrees with intuition obtained from the trivial limits $\theta_{12}\to0$~or~$\pi/2$, when $\nu_{\mu}\leftrightarrow\nu_{\mu}$ oscillations can depend only on $\Delta m^2_{23}$ or $\Delta m^2_{13}$, respectively.} Under theses circumstances, it is impossible to tell one mass hierarchy from the other.

Another easy-to-describe limiting case is $\Delta^m_{12}L\ll 1$. Under these circumstances, we can use the results of Sec.~\ref{sec:vacuum} to determine the value of $x$ that renders $P_{\mu\mu}^+=P_{\mu\mu}^-$ without too much effort. We start by computing the auxiliary quantity $y^{m}$, defined by $\Delta_{13}^{-,m}=-\Delta_{13}^{+,m}+Y^m$ ($Y^m=y^m/2E$), which renders $P_{\mu\mu}(\Delta_{13}^{+,m})=P_{\mu\mu}(\Delta_{13}^{-,m})$. The definition of $y^m$ is identical to the one for $x$ (see Sec.~\ref{sec:vacuum}) after all parameters are replaced by effective ones. While $y^m$ has no simple physical interpretation, it can be related to $x$ in a simple way and we can easily solve for it.

Because the expression for $P_{\mu\mu}$ in matter is the same as the one in vacuum with the vacuum mixing parameters replaced by effective ones, we can simply read off the solution from Eq.~(\ref{eq:xaa}): $Y^m=2\cos^2\theta_{12}^m\Delta^m_{12}=\Delta_{12}^m-A+\Delta_{12}\cos2\theta_{12}$. We can relate $y^m$ to $x$:
\begin{eqnarray}
\frac{-A-\Delta_{12}+2\Delta_{13}^-+\Delta_{12}^m}{2}&=&\frac{A+\Delta_{12}-2\Delta^+_{13}-\Delta_{12}^m}{2}+Y^m, \\
\Delta^-_{13}&=&-\Delta_{13}^++A+\Delta_{12}-\Delta_{12}^m+Y^m, \\
\Delta^-_{13}&=&-\Delta_{13}^++X,
\end{eqnarray}
where in the last step we simply repeat the definition of $x$. Hence, 
\begin{equation}
X=\frac{x}{2E}=Y^m+A+\Delta_{12}-\Delta^m_{12}, \label{xm_to_x}
\end{equation}
and, since  $Y^m=\Delta_{12}^m-A+\Delta_{12}\cos2\theta_{12}$, $x=2\Delta m^2_{12}\cos^2\theta_{12}$. This result agrees with the vacuum value of $x$ obtained in the case $\Delta_{12}L\ll 1$, as expected. As is well known, for two flavor oscillations $P_{\alpha\beta}^{\rm matter}=P_{\alpha\beta}^{\rm vac}$ in the limit $\Delta^mL\ll 1$ and it also turns out that here, in the limit $\Delta^m_{12}L\ll 1$, $P_{\alpha\beta}^{\rm matter}=P_{\alpha\beta}^{\rm vac}$ for all $\alpha,\beta=e,\mu,\tau$.

The two limiting cases discussed above reveal that, in order to break the normal--inverted degeneracy in $\nu_{\mu}\leftrightarrow\nu_{\mu}$ oscillations, it is necessary to probe $P_{\mu\mu}$ for $L$ and $E$ values such that $\Delta_{12}L\gtrsim 1$,  independent of whether there are matter effects. Numerically,
\begin{equation}
\Delta_{12}L=1.2\left(\frac{\Delta m^2_{12}}{8\times 10^{-5}~\rm eV^2}\right)\left(\frac{0.5~\rm GeV}{E}\right)\left(\frac{L}{3000~\rm km}\right).
\end{equation}
Furthermore, one needs to steer clear of very strong matter effects, which translates into $E<$~few GeV for Earth-based sources and detectors and very long baselines. Similar neutrino energies and baselines have, of course, also been identified as the optimal conditions to study matter effects in the 12 sector \cite{winter}. 

We proceed to discuss how to determine the mass hierarchy following the strategy discussed in the previous subsection. We compute the value of $x$ that renders $P_{\mu\mu}^+=P_{\mu\mu}^-$ at a fixed value of $L/E$. Following our previous discussion, it is straightforward to obtain 
\begin{equation}
x=2EA+\Delta m^2_{12}+\frac{4E}{L}\arctan\left(\cos2\theta^m_{12}\tan\left(\frac{\Delta^m_{12}L}{2}\right)\right)+4n\pi\frac{E}{L},~{\rm integer}~n,
\label{eq:x(L,E)}
\end{equation}
using Eqs.~(\ref{eq:x(L/E)}) and (\ref{xm_to_x}). 

\begin{figure}
{\centering
\epsfig{figure=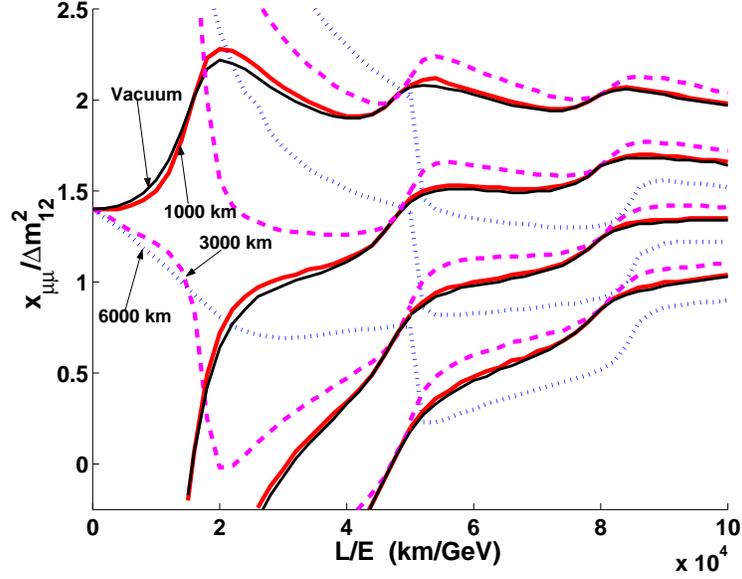,width=0.55\textwidth}
}
\caption{Values of $x_{\mu\mu}/\Delta m^2_{12}$ that render $P_{\mu\mu}^+(L,E)=P_{\mu\mu}^-(L,E)$ in constant matter density $\rho$, as a function of $L/E$, for $\Delta m^2_{12}=7.9\times 10^{-5}$~eV$^2$, $\sin^2\theta_{12}=0.3$, and  $\rho=0$ (vacuum oscillations, solid black line), $L=1000$~km, $\rho=1.50$g/cm$^3$ (solid red line), $L=3000$~km, $\rho=3.14$g/cm$^3$ (dashed pink line), and $L=6000$~km, $\rho=3.73$g/cm$^3$ (dotted blue line). See text for details.}
\label{Fig:xmatter}
\end{figure}

Figure~\ref{Fig:xmatter} depicts $x$ values that render $P_{\mu\mu}^+=P_{\mu\mu}^-$ in matter as a function of $L/E$ for different baselines, assuming $\Delta m^2_{12}=7.9\times 10^{-5}$~eV$^2$ and $\sin^2\theta_{12}=0.3$. We choose matter densities equal to the average matter density value along the neutrino path, assuming that the neutrinos are produced and detected near the Earth's surface, and using the preliminary Earth model \cite{PREM}. We extend all curves to very large values of $L/E$, even if these correspond to experimentally unexaminable values of the neutrino energy ($E\lesssim 100$~MeV). In order to avoid an overly busy plot, we only display a few values of $n$ (see Eq.~(\ref{eq:x(L,E)})) for each $L$ and $E$.

As advertised, significant effects are expected for large values of $L$ ($L> 1000$~km) and small energy values ($L/E\gtrsim 5000$~km/GeV). In this region, not only is $x\neq x(L/E\to 0)$, but also $x^{\rm matter}\neq x^{\rm vac}$, as long as $L/E$ is not too large.  Very large $L/E$ values correspond to very small energy values and $\Delta_{12}\gg A$, in which case matter effects become unobservable. This behavior can be seen in Fig.~\ref{Fig:xmatter} for $L=3000$~km and $L/E>\rm several\times10^{4}$~km/GeV. This regime cannot be observed in practice, as  $\Delta_{12}\gg A$ implies $E<100$~MeV for typical Earth matter densities.

Fig.~\ref{Fig:xmatter} reveals several distinct ways of determining the mass hierarchy. As briefly discussed in the previous subsection, a ``broad-band'' experiment with, say, a 6000~km baseline capable of precisely measuring the atmospheric mass-squared difference for energy values around, say, $E=1.2$~GeV ($L/E=0.5\times 10^{4}$~km/GeV) and $E=300$~MeV ($L/E=2\times 10^{4}$~km/GeV) should be able to determine the mass hierarchy as long as $\Delta(\Delta m^2_{13})\lesssim 0.5\Delta m^2_{12}$. By ``measuring the atmospheric mass-squared difference at some particular energy value,'' we mean the following. Select a narrow window around a particular $L/E$ value (say $L/E\pm 0.1L/E$). Since we need values of $L/E=O(10^4~\rm km/GeV)$, we should find at least a few atmospheric oscillations inside our $L/E$ window. Numerically, 
\begin{equation}
\frac{\Delta_{13}L}{2}=10.1\pi\left(\frac{\Delta m^2_{13}}{2.5\times 10^{-3}~\rm eV^2}\right)\left(\frac{L/E}{10^{4}~\rm km/GeV}\right),
\end{equation} 
and it is clear that a very big challenge for ``broad-band'' neutrino experiments is to make sure that atmospheric effects do not ``average out'' --- one needs to be able to ``see'' not the first and second oscillation maxima, but the tenth and the eleventh! This implies, among other things, excellent energy resolution (and angular resolution in the case of atmospheric neutrinos), and precise knowledge of the neutrino beam. A detailed quantitative study of the experimental requirements is clearly outside the scope and philosophy of this paper. 

One can also determine the mass hierarchy by performing two different experiments at the same value of $L/E$ but different $L$. For example, a 3000~km baseline experiment performed at $L/E$ around $2\times 10^{4}$ would find $x\sim 1.4\Delta m^2_{12}$ (among other values), while a different experiment performed around the same value of $L/E$ but, say,  $L=6000$~km, would encounter $x=0.7\Delta m^2_{12}$. If both experiments can determine the atmospheric mass-squared difference with an uncertainty smaller than $|1.4-0.7|\Delta m^2_{12}$, they should be able to unveil whether the neutrino mass hierarchy is normal or inverted.

\subsection{Comparing $P(\nu_{\mu}\to\nu_{\mu})$ to $P(\bar{\nu}_{\mu}\to\bar{\nu}_{\mu})$}

In the previous subsections, we described how different $\nu_{\mu}$ disappearance searches can reveal the neutrino mass ordering, including the fact that neutrinos will often traverse a significant amount of matter on their way to the detector. When this is the case, it is well known that $P_{\mu\mu}$ will differ from $P_{\bar{\mu}\bar{\mu}}$ because matter effects violate CP-invariance\footnote{For $|U_{e3}|=0$, there are no intrinsic CP-invariance violating effects in neutrino$\leftrightarrow$neutrino oscillations.} (and CPT-invariance). It seems, therefore, plausible that one should be able to determine the mass hierarchy by comparing muon neutrino and muon antineutrino disappearance.

\begin{figure}
{\centering
\epsfig{figure=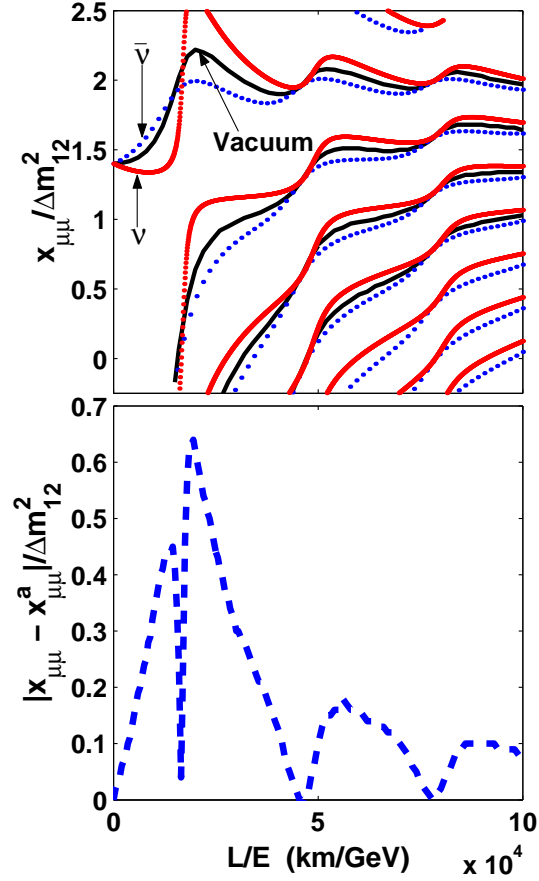,width=0.4\textwidth}
}
\caption{(top panel) Values of $x_{\mu\mu}/\Delta m^2_{12}$ and $x_{\bar{\mu}\bar{\mu}}/\Delta m^2_{12}$ that render, respectively,  $P_{\mu\mu}^+(L,E)=P_{\mu\mu}^-(L,E)$ (solid red line) and $P_{\bar{\mu}\bar{\mu}}^+(L,E)=P_{\bar{\mu}\bar{\mu}}^-(L,E)$ (dotted blue line), and (bottom panel) the minimum value of $|x_{\mu\mu}-x_{\bar{\mu}\bar{\mu}}|/\Delta m^2_{12}$, as a function of $L/E$. $L=2000$~km, $\rho=2.83$g/cm$^3$, $\Delta m^2_{12}=7.9\times 10^{-5}$~eV$^2$, and $\sin^2\theta_{12}=0.3$. For comparison, the equivalent $x_{\mu\mu}=x_{\bar{\mu}\bar{\mu}}$ values in vacuum are also depicted in the top panel (solid black line). }
\label{Fig:nu_anti_nu}
\end{figure}

We expect  $P_{\bar{\mu}\bar{\mu}}\neq P_{\mu\mu}$ only for $\Delta_{12}L\gtrsim 1$ and $\Delta_{12}/A\sim 1$ given that (i) in the limit $\Delta_{12}\gg A$ or $\Delta_{12}L\ll 1$, $P_{\mu\mu}=P_{\mu\mu}^{\rm vac}$, (ii) in the limit $A\gg \Delta_{12}$, $P_{\mu\mu}=P_{\bar{\mu}\bar{\mu}}$, both given by Eq.~(\ref{Pmumu_largeA}). All of these features can be observed in Fig.~\ref{Fig:nu_anti_nu}(top), which depicts the values of $x$ that render  $P_{\mu\mu}^+= P_{\mu\mu}^-$  and those that render  $P_{\bar{\mu}\bar{\mu}}^+= P^-_{\bar{\mu}\bar{\mu}}$ as a function of $L/E$, for $L=2000$~km, $\Delta m^2_{12}=7.9\times 10^{-5}$~eV$^2$, $\sin^2\theta_{12}=0.3$, and $\rho=2.83$~g/cm$^3$.

Fig.~\ref{Fig:nu_anti_nu}(bottom) reveals that the comparison of results obtained with a muon neutrino and antineutrino beam performed at $L=2000$~km and, say, $E=200$~MeV ($L/E=1\times 10^{4}$~km/GeV) will determine the neutrino mass-hierarchy as long as $\Delta(\Delta m^2_{13})\lesssim0.5\Delta m^2_{12}$.

\setcounter{footnote}{0}
\setcounter{equation}{0}
\section{Summary and Conclusions}
\label{sec:end}

One of the challenges of next-generation neutrino experiments is to establish the neutrino mass hierarchy: is the relatively isolated $\nu_3$ state (much) heavier than $\nu_1$ and $\nu_2$, or is it (much) lighter (see Fig.~\ref{normal_inverted})? While we currently do not know the answer to this question, we do know that these two possibilities --- a normal or an inverted mass hierarchy --- are dramatically different.\footnote{We discount the extraordinary possibility that all three neutrino masses are quasi-degenerate. Whether the neutrino masses are all quasi-degenerate cannot be probed by oscillation experiments, and is already somewhat constrained by searches for neutrinoless double beta decay and indirect studies of the relic neutrino background.} In the case of a normal hierarchy, the neutrino masses can be ordered just like the quark and charged lepton masses --- $m_{1}\ll m_{2} \ll m_{3}$. In the case of an inverted mass hierarchy, the neutrino masses are ordered in a most unusual way --- $m_3\ll m_{1,2}$ and $m_2-m_1\ll (m_2+m_1)$, {\it i.e.}, two of the neutrino mass eigenstates are quasi-degenerate. We are yet to encounter two fundamental particles with quasi-degenerate masses. It is not surprising, therefore, that the determination of the neutrino mass hierarchy has been identified as one of the key discriminators of neutrino mass models \cite{neutrino_theory}.
 
 In the literature, studies of matter--affected 13 oscillations have been singled out as the most realistic means of establishing the character of the neutrino mass hierarchy. It is interesting that these studies do not rely on properties unique to three (or more) neutrino flavor oscillations, and significantly mirror the (successful) path taken in order to determine the ``solar'' neutrino mass hierarchy. Such studies, briefly described in Sec.~\ref{sec:traditional}, require  $|U_{e3}|$ to be large, and utterly fail if the limit  $|U_{e3}|\to0$ applies. 
 
In this paper, we have concentrated on probes of the neutrino mass hierarchy that do not rely on neutrino oscillations in matter or on $|U_{e3}|$ being ``large.'' Instead, they rely heavily on the fact that, given that there are at least three neutrinos, $P_{\alpha\beta}$ will depend on at least two different oscillation frequencies. 

First, we show that, in principle, vacuum oscillations are capable of distinguishing an inverted from a normal mass hierarchy --- for a given normal mass hierarchy, there is no inverted mass hierarchy capable of producing the same $P_{\alpha\beta}(L/E)$. We also argue that only neutrino oscillation experiments that are simultaneously sensitive to $\Delta m^2_{13}$ and $\Delta m^2_{12}$ effects can ``see'' the mass hierarchy. Hence, we need experiments capable of measuring $|\Delta m^2_{13}|$ with precision $\Delta(\Delta m^2_{13})$ smaller than $\Delta m^2_{12}$.

Second, we argue that experiments performed at $L/E$ values such that $\Delta_{12} L\ll 1$ suffer from a normal--inverted degeneracy, which can only be broken if more experimental information is available. The same is true of  ``narrow-band-beam'' experiments which probe ``around'' a fixed value of $L$ and $E$, regardless of the magnitude of $\Delta_{12}L$. 
 
 If $|U_{e3}|\neq 0$, the neutrino mass hierarchy can be unveiled by combining searches for $\nu_{\mu}$ disappearance with searches for $\bar{\nu}_e$ (or $\nu_e$) disappearance, as long as both experiments are capable of measuring $|\Delta m^2_{13}|$ with precision $\Delta(\Delta m^2_{13})$ smaller than $\Delta m^2_{12}$. It seems unlikely that this precise a measurement of $\Delta m^2_{13}$ via $\bar{\nu}_e$ disappearance can be performed by reactor neutrino experiments. The KamLAND reactor experiment has measured $\Delta m^2_{12}$ with precision $\Delta(\Delta m^2_{12})\simeq 8\%$, which we take as an optimistic estimate of the precision with which a next generation reactor experiment can measure $\Delta m^2_{13}$ if $|U_{e3}|^2\gtrsim 10^{-2}$. 
  
 In the limit $|U_{e3}|\to0$, any information regarding the neutrino mass hierarchy (if one sticks to neutrino oscillation probes) has to come from searches of $\nu_{\mu}$ or $\bar{\nu}_{\mu}$ disappearance at distinct baselines and neutrino energies. If these experiments are to be performed on the Earth, at least one of them will require a very long baseline ($L/E\gtrsim 5000$~km/GeV) and relatively small neutrino energies ($E\lesssim 1$~GeV) in order to guarantee that $\Delta_{12}L\gtrsim 1$ and to avoid very strong matter effects ($\sqrt{2}G_FN_e\gg\Delta_{12}$), respectively.  Different potential scenarios include comparing $\nu_{\mu}$ disappearance searches at short $(\Delta _{12}L\ll 1)$ and long ($\Delta_{12}L\sim 1$) baselines, comparing searches for $\nu_{\mu}$ disappearance at different $L$ but the same value of $L/E$, or comparing $\nu_{\mu}$ and $\bar{\nu}_{\mu}$ disappearance in the presence of nontrivial matter effects ($\sqrt{2}G_FN_e\sim\Delta_{12}$). Note that ``broad-band'' accelerator experiments or studies of atmospheric neutrinos are potentially capable of performing all of these comparisons without the aid of other experimental setups --- one can interpret a broad-band/atmospheric experiment as a collection of multiple experiments at different values of $L/E$. 
 
 Instead of concentrating on the feasibility of any of the setups summarized above, we chose to identify the conditions that need to be met in order for one to successfully distinguish an inverted mass hierarchy from a normal one. We find that, irrespective of the scenario, we require multiple experiments capable of measuring $|\Delta m^2_{13}|$ at the few percent level ($6.4\%>\Delta m^2_{12}/|\Delta m^2_{13}|>2.2\%$, taking three sigma bounds for mass-squared differences \cite{global_anal}). Whether this can be achieved in practice will be left for detailed experimental studies. We would, however, like to summarize the progress that can be expected in the foreseeable future.\footnote{ Note that none of the proposed experiments currently under serious consideration meet the requirements for revealing the neutrino mass hierarchy spelled out above: either the baseline is too short, the energies too high, or the energy resolution insufficient to probe 13 oscillations at low energies.} Today, $\Delta(\Delta m^2_{13})\sim 25\%$ \cite{global_anal}. In the near future, the MINOS experiment expects to measure the atmospheric mass-squared difference with $\Delta(\Delta m^2_{13})\sim 10\%$ \cite{MINOS}, while next-generation accelerator experiments aim at $\Delta(\Delta m^2_{13})\sim 1\%$ \cite{lbl_pheno}. In order to achieve such precise numbers, it is clear that improved knowledge of the neutrino beams and the neutrino--target cross sections are needed.
 
 We wish to conclude by reemphasizing that neutrino oscillation experiments can, in principle, address a most relevant and fundamental question in high energy physics --- what is the character of the neutrino mass hierarchy? This capability does not depend on whether the oscillations occur in matter or on whether $|U_{e3}|$ is nonzero. If, however, $|U_{e3}|$ is sufficiently small, our study here shows that unveiling the mass hierarchy through oscillations may prove very hard. Detailed studies are required in order to determine how hard.

\section*{Acknowledgments}

AdG and BK are grateful to Hamish Robertson for a conversation that ultimately lead to this paper, and to Stephen Parke for enlightening discussions. BK would like to thank Gabriela Barenboim for the hospitality of the Universitat de Val\`encia, where part of this work was completed. Fermilab is supported by the US Department of Energy Contract DE-AC02-76CHO3000. The work of AdG is sponsored in part by the US Department of Energy Contract DE-FG02-91ER40684.

\appendix
\setcounter{footnote}{0}
\setcounter{equation}{0}
\section{Analytic Expression for the Oscillation Probabilities in Constant Matter Density and Vanishing $|U_{e3}|$}
\label{app:tech}

Simple analytic expressions for constant matter density neutrino
oscillation probabilities can be obtained in the case $|U_{e3}|=0$. These are presented in this appendix. Other very useful approximate expressions, including $O(|U_{e3}|)$ corrections to the results derived below, can be found, for example, in \cite{P_expressions}.

Neutrino oscillations in matter are described by the following set of differential equations:
\begin{equation}
\imath\frac{d\vec{\nu}}{dL}=\textbf{M}\vec{\nu},
\label{diff_equations}
\end{equation}
where
\begin{equation}\vec{\nu} = \left(\begin{array}{ccc} \nu_{e} \\ \nu_{\mu} \\
\nu_{\tau} \end{array}\right)\end{equation}
is a vector of neutrino flavor eigenstates and \begin{equation}\textbf{M} =
\mathbf{U}\left(\begin{array}{ccc}
0 & 0 & 0\\
0 & \Delta_{12} & 0\\
0 & 0 & \Delta_{13}\\
\end{array}\right)\mathbf{U^{\dag}}+\left(\begin{array}{ccc}
A & 0 & 0\\
0 & 0 & 0\\
0 & 0 & 0\\
\end{array}\right).
\end{equation} 
The $\Delta_{ij} = \frac{\Delta m_{ij}^{2}}{2E}$ parameters
contain all information regarding the neutrino mass-squared
differences, and describe the energy dependence of the
system.  The forward scattering of electron neutrinos in matter is
accounted for by $A = \sqrt{2}G_{F}N_{e}$, where $G_{F}$ is the
Fermi coupling constant, and $N_{e}$ is the electron number density
of the medium. As a function of the mass density $\rho$, $A$ can be expressed
numerically as $A = 3.78\times 10^{-5}
eV^{2}\frac{\rho}{\rm g/cm^3}$ after we fix the electron to baryon fraction $Y=1/2$. $\mathbf{U}$ is a unitary mixing
matrix.  In the standard parametrization \cite{pdg} with $\theta_{13}$ set to zero, and ignoring potential 
Majorana phases, which are not relevant here,
\begin{equation}
\mathbf{U} = \left(\begin{array}{ccc} \cos\theta_{12}&\sin\theta_{12}&0\\
-\sin\theta_{12} \cos\theta_{23} & \cos\theta_{12}\cos\theta_{23} &
\sin\theta_{23}\\
\sin\theta_{12}\sin\theta_{23} & -\cos\theta_{12}\sin\theta_{23} &
\cos\theta_{23}\\
\end{array}\right) \equiv \left(\vec{u}_1~\vec{u}_2~\vec{u}_3\right),
\label{eq:MNS}
\end{equation}
where $\vec{u}_i$ are column vectors. Since $\mathbf{U}$ is unitary, $\vec{u}_i^{\dagger}{\vec{u}}_j=\delta_{ij}$, which means that $\vec{u}_i$ form an orthonormal basis for the space of three component vectors.

The solutions to the three coupled differential equations in Eq.~(\ref{diff_equations}) can be expressed in terms of the eigenvalues $\lambda_{i}$ and eigenvectors $\vec{v}_{i}$ of \textbf{M}, $i=1,2,3$.  Even though \textbf{M} is a three by three matrix, $\lambda_{i}$ and $\vec{v_{i}}$ may be found very easily. 

\textbf{M}, when $\mathbf{U}$ is given by Eq.~(\ref{eq:MNS}), may be expressed in the following convenient form:
\begin{equation}
\mathbf{M}=\Delta_{12}\vec{u}_2\vec{u}_2^{\dagger}+\Delta_{13}\vec{u}_3\vec{u}_3^{\dagger}+A(\cos\theta_{12}\vec{u}_1+\sin\theta_{12}\vec{u}_2)(\cos\theta_{12}\vec{u}_1+\sin\theta_{12}\vec{u}_2)^{\dagger}.
\end{equation}
It is clear that one of the eigenvectors of $\mathbf{M}$ is $\vec{u}_3$: $\mathbf{M}\vec{u}_3=\Delta_{13}\vec{u}_3$. Furthermore, since $\mathbf{M}$ is Hermitian, the other two eigenvectors are orthogonal to $\vec{u}_3$, which means they can be expressed as orthogonal linear combinations of $\vec{u}_1$ and $\vec{u}_2$. The problem of finding the three eigenvalues and eigenvectors of a three by three matrix is now reduced to finding two eigenvectors in a two by two effective matrix, which is very easily solved. Indeed, the problem reduces to the well known case of two-flavor neutrino oscillations in constant matter, and its solution is very well known \cite{TASI}. 

Hence, the resulting oscillation probabilities are conveniently expressed in terms of
effective mixing angles and frequencies, which are simply related to their vacuum
counterparts.
\begin{eqnarray}
\Delta_{12}^{m} &=& \sqrt{A^{2} - 2A\Delta_{12}\cos2\theta_{12} +
\Delta_{12}^{2}}, \\
\Delta_{13}^{m} &=& \frac{-A - \Delta_{12} + 2\Delta_{13} + \Delta_{12}^m}{2},\\
\Delta_{23}^{m} &=& \frac{-A - \Delta_{12} + 2\Delta_{13} - \Delta_{12}^m}{2}, \\
\sin2\theta_{12}^{m} &=&\frac{\Delta_{12}}{\Delta_{12}^m}\sin2\theta_{12},\\
\cos2\theta_{12}^{m} &=&\frac{\Delta_{12}\cos2\theta_{12}-A}{\Delta_{12}^m}, \\
\theta_{23}^m&=&\theta_{23},
\end{eqnarray}
$\Delta_{13}^m-\Delta_{23}^m=\Delta_{12}^m$, and $\Delta_{12}^m$ reduces to
$\Delta_{12}$ in the vacuum limit $\Delta_{12}\gg A$. 

The oscillation probabilities coincide exactly with 
the corresponding vacuum expressions once the vacuum oscillation parameters are replaced by the corresponding
effective ones. We explicitly list all oscillation probabilities in constant matter below, keeping in mind that, due to T-invariance,\footnote{When $|U_{e3}|=0$, there are no T-invariance violating effects in neutrino oscillations in vacuum. Furthermore, a constant matter potential does not induce T-invariance violating effects in neutrino oscillations in matter. Matter-induced CP-invariance violation is, of course, present.} $P_{\alpha\beta} = P_{\beta\alpha}$, $\alpha,\beta=e,\mu,\tau$. The oscillation probabilities of antineutrinos are given by the same expressions, with $A\to -A$.
\begin{eqnarray}
P_{ee} &=& 1 -
\sin^{2}2\theta_{12}^{m}\sin^{2}\frac{L\Delta_{12}^{m}}{2}, \\
P_{e\mu} &=&
\sin^{2}2\theta_{12}^{m}\cos^{2}\theta^m_{23}\sin^{2}\frac{L\Delta_{12}^{m}}{2},\\
P_{e\tau} &=&
\sin^{2}2\theta_{12}^{m}\sin^{2}\theta^m_{23}\sin^{2}\frac{L\Delta_{12}^{m}}{2}. 
\end{eqnarray}
As expected, $P_{e\alpha}$ ($\alpha=e,\mu,\tau$) does not depend on $\Delta m^2_{13}$, and hence does not care whether the neutrino mass hierarchy is normal or inverted. This is not the case of $P_{\mu\mu,\mu\tau,\tau\tau}$, which are given by
\begin{eqnarray}
P_{\mu\mu} &=& 1 -
\sin^{2}2\theta_{12}^{m}\cos^{4}\theta_{23}\sin^{2}\frac{L\Delta_{12}^{m}}{2}
-\sin^{2}2\theta^m_{23}\left(\sin^{2}\theta_{12}^{m}\sin^{2}\frac{L\Delta_{13}^{m}}{2}
+ \cos^{2}\theta_{12}^{m}\sin^{2}\frac{L\Delta_{23}^{m}}{2}\right), \\
P_{\mu\tau} &=&
-\frac{\sin^{2}2\theta_{12}^{m}\sin^{2}2\theta_{23}}{4}\sin^{2}\frac{L\Delta_{12}^{m}}{2}
+\sin^{2}2\theta^m_{23}\left(\sin^{2}\theta_{12}^{m}\sin^{2}\frac{L\Delta_{13}^{m}}{2}
+ \cos^{2}\theta_{12}^{m}\sin^{2}\frac{L\Delta_{23}^{m}}{2}\right),\\
P_{\tau\tau} &=& 1 -
\sin^{2}2\theta_{12}^{m}\sin^{4}\theta_{23}\sin^{2}\frac{L\Delta_{12}^{m}}{2}
-\sin^{2}2\theta^m_{23}\left(\sin^{2}\theta_{12}^{m}\sin^{2}\frac{L\Delta_{13}^{m}}{2}
+ \cos^{2}\theta_{12}^{m}\sin^{2}\frac{L\Delta_{23}^{m}}{2}\right).
\end{eqnarray}
The terms of $P_{\mu\mu,\mu\tau,\tau\tau}$ that depend on $\Delta m^2_{13}$ are identical, up to an overall sign. Hence, $|P^+-P^-|$ is the same for all non-electron oscillation probabilities. Any normal--inverted degeneracy observed in a study of $\nu_{\mu}\to\nu_{\mu}$ will be present in exactly the same form in a similar study of $\nu_{\mu}\to\nu_{\tau}$ or $\nu_{\tau}\to\nu_{\tau}$.

 \end{document}